\documentclass[aps,prb,twocolumn]{revtex4}
\usepackage[dvips]{epsfig}
\usepackage{psfrag}
\usepackage{amssymb,amsmath}
\usepackage[dvips]{graphicx}
\usepackage[usenames,dvipsnames]{color}
\usepackage{float}

\begin{document}

\title{Quantum phase diffusion in ac-driven
  superconducting atomic point contacts}

\author{M. F. Carusela}
\email{flor@ungs.edu.ar}
\affiliation{Instituto de Ciencias, Universidad Nacional de General Sarmiento,Buenos Aires, Argentina, \\ Conicet,Argentina}
\author{J. Ankerhold}
\affiliation{Institut f\"ur theoretische Physik, Universit\"at Ulm,
Albert-Einstein-Allee 11, D-89069 Ulm, Germany}

\date{\today}

\begin{abstract}
The impact of quantum fluctuations on the phase diffusion in resistively shunted superconducting quantum points subject to an external ac-voltage is studied. Based on an extension of the classical Smoluchowski equation to the quantum regime, numerical results for fractional and integer Shapiro resonances are investigated to reveal characteristic features of quantum effects. It is shown that typically in the quantum regime the broadening of resonances cannot be described simply by an effective temperature.
\end{abstract}


\maketitle

\section{Introduction}

In recent years there has been a notable advance in the understanding of electronic transport through superconducting nanosystems. In particular, the development of fabrication techniques such as scanning tunneling microscopy, break-junction and lithographic methodologies \cite{review} have allowed to study atomic-size metallic contacts as ideal systems to test fundamental properties of charge transfer through superconducting weak links.
These achievements have not only deepened our understanding of subgap structures in the current-voltage characteristics but also revealed even microscopic details of the contact such as transmission coefficients \cite{muller,Cron}.  Theoretical predictions based on Greens-function or mean-field approaches have been confirmed in various experiments \cite{averin,cuevas}.  In this context, set-ups where atomic-size tunnel junctions are subject to both dc- and ac-voltages give access to the intimate relation between phase dynamics, driving, and dissipation leading to pronounced Shapiro resonances of integer and fractional order.

Similar to conventional weak links, atomic point contacts are characterized  by two energy scales \cite{agrait}, namely,  the coupling energy between the superconducting domains (Josephson energy) and the charging energy of the junction. The competition between these two scales is crucially influenced by the electromagnetic environment so that  a realistic modeling of charge transport across the contact must necessarily incorporate its embedding in
an actual circuit. Accordingly, the dynamics of the superconducting phase difference as the only relevant degree of freedom exhibits a diffusive motion subject to noise which, as it is well-known from the physics of Josephson junctions \cite{barone}, can be visualized as the Brownian motion of a fictitious particle.
It turned out that for atomic point contacts this phase dynamics occurs in an  overdamped regime. The corresponding classical frame is provided by the Smoluchowski equation which has already been the starting point for calculations of current-voltage characteristics of Josephson junctions in low impedance environments \cite{IZ,AH}. There, only the Josephson energy remains as relevant parameter while charging effects related to inertia drop out. The impact of quantum fluctuations have been attacked within a time-dependent perturbation theory in \cite{grabert} where the whole range from
coherent to incoherent Cooper pair transfer in the domain of Coulomb blockade could be captured. Later, a generalization of the Smoluchowski approach to the quantum regime (Quantum Smoluchowski) developed by one of us (JA) and co-workers \cite{qmsmolu1,qmsmolu2,Anker,Anker-libro} has allowed to derive the same physics in a very elegant manner and in close analogy to the classical description. Quantum noise has been shown to be inevitably associated with charging effects according to the uncertainty principle, thus physically ruling the changeover toward Coulomb blockade dominated transport.

 The motivation for the present work is two-fold. On the one hand, it is based on experiments conducted with atomic point contacts in the last years in the Quantronics group \cite{Chauvintesis,stein} and on the other hand on a corresponding description in terms of the classical Smoluchowski approach \cite{dupret}. While experimental results with ac-driven junctions followed theoretical predictions for the structure of Shapiro resonances, substantial discrepancies appeared for their heights and widths. A plausible explanation has been the presence of residual spurious noise sources which lead to an effective temperature at the contact different from the actual base temperature. Here, we analyze if and if yes to what extent quantum noise must also be incorporated into this picture. Eventually, this may open the door to unambiguously characterize quantum effects in overdamped systems at low temperatures.

The paper is organized as follows. In Sec.~\ref{model} we present our generalization of the RSJ model for contacts of arbitrary transmission in the presence of microwave radiation and in the presence of quantum fluctuations. In Sec.~\ref{numerics} the numerical method to solve the generalized quantum Smoluchowski equation is outlined together with the relevant scales and approximations. Section \ref{results} discusses results for the $I-V$ characteristics and fractional Shapiro resonances, before in Sec.~\ref{effective} the question whether quantum noise can be captured by an effective temperature is addressed. Section \ref{conclusions} is devoted to discussion and conclusions on future experimental realizations.

\section{The Model}\label{model}

We model a superconducting tunnel junction using an equivalent circuit with so-called lumped circuit parameters that includes both the effect of  dissipative sources and the distributed capacity. For weak links the standard resistively and capacitively shunted junction (RCSJ) model captures the essential physics even in presence of an external ac-voltage \cite{barone}. The equivalent circuit, shown in Fig.~\ref{fig:circuit}, is formed by a contact with a resistance $R$, a capacitance $C$, and biased by a dc-current $I_{b}$.
The superconducting phase difference across the junction  denoted by $\theta$ is the only relevant degree of freedom with a current-phase relation $I(\theta)$. Energy dissipation in the contact is accompanied by Johnson-Nyquist noise $I_n(t)$ thus the current conservation relation gets:
\begin{figure}[ht]
\epsfig{file=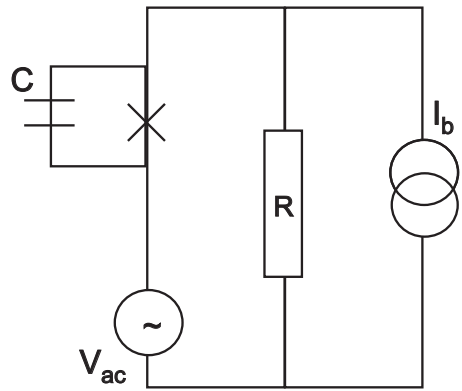, width=6cm}
\caption{Lumped circuit model of a superconducting weak link with capacitance $C$ and within a resistive environment $R$ subject to a dc-bias current $I_b$ and an ac-voltage.}
\label{fig:circuit}
\end{figure}
\begin{equation}
I_{b} = C \frac{dV}{dt} + I(\theta)+ \frac{V_{\rm tot}(t)}{R} + I_n(t)\, .
\label{eq:lan1}
\end{equation}
Here, the first term on the RHS is the displacement and the third term is the dissipative current.
 Further, $V_{\rm tot}(t)=V(t)+V_{ac} \cos(\omega t)$ with voltage $V$ across the contact and $V_{ac}$ being an additional ac-voltage induced by a microwave field.
  According to $V=\Phi_0 \dot{\theta}$ ($\Phi_0=\hbar/2e$) this equation is in fact an equation of motion for the phase, i.e.,
 \begin{equation}
\frac{V_{ac}(t)}{R}+I_{b}=\Phi_0 C \ddot{\theta}(t)+ \frac{\Phi_0}{R}\dot{\theta}(t) +I(\theta)+ I_n(t)\, ,
\label{eq:lan1b}
\end{equation}
which is equivalent to the Brownian motion of a fictitious particle with mass $M \equiv \Phi_0^2 C$, friction constant  $\eta \equiv 1/RC$ and classical noise force $Z_{Cl}(t) \equiv -R I_n(t)/\Phi_0$ in the potential
 \begin{equation}
U(\theta,t) = \Phi_0 \int_{0}^{\theta} I(\theta) d\theta -\Phi_0 \theta [I_{b}-\frac{V_{ac}}{R} cos(\omega t)]\, .
\label{eq:pot}
\end{equation}
The noise force has zero mean and obeys
\begin{equation}
\langle Z_{Cl}(t)Z_{Cl}(t^{\prime })\rangle=\frac{2D_{Cl}}{\eta M}\, \delta (t-t^{\prime }).
\end{equation}
with $D_{Cl}=k_{B}T\equiv 1/\beta$.

The regime where the capacitance is negligible thus corresponds to the strong friction domain (Smoluchowski regime) in the mechanical analog \cite{barone}. Previous work has shown that this is indeed the range where phase diffusion in superconducting atomic point contacts happens to occur \cite{Chauvintesis,dupret}. Strong friction considerably simplifies the description and for the situation with zero driving even allows for analytical results. It is then often convenient to switch from the classical Langevin equation corresponding to (\ref{eq:lan1b}), i.e.,
\begin{equation}
\dot{\theta} =  \frac{1}{\eta M}\frac{dU(\theta,t)}{d\theta} + Z_{Cl}(t)
\label{eq:lan2}
\end{equation}
to an equation of motion for the probability distribution $P(\theta, t)$, namely,
\begin{equation}
\frac{\partial P(\theta,t)}{\partial t} = \frac{1}{\eta M}\frac{\partial}{\partial \theta}\left[-\frac{\partial U(\theta,t)}{\partial \theta} +D_{Cl} \frac{\partial}{\partial \theta}\right] P(\theta,t)\, .
\label{eq:smol}
\end{equation}

As first pointed out in \cite{qmsmolu1}, this classical Smoluchowski equation (SE) can be generalized to the low temperature domain where quantum fluctuations become substantial \cite{qmsmolu2}. The quantum Smoluchowski equation (QSE) has been studied since then in a variety of applications \cite{Anker-libro} including particularly an extension of the classical Ivanchenko-Zil'berman theory for Josephson junctions in low impedance environments \cite{IZ,AH}. There, quantum fluctuations are related to charging effects and reveal signatures of Coulomb blockade physics.
The QSE follows from its classical counterpart by replacing $D_{Cl}\ \to \ D_{Q}(\theta)$ with the position dependent quantum diffusion coefficient
\begin{equation}
 D_{Q}(\theta)=\dfrac{k_{B}T}{1-\Lambda \beta U''(\theta)}
\label{eq:qsmol}
\end{equation}
with a friction and temperature dependent function
\begin{equation}
\Lambda=2\rho\left[c+\frac{2\pi^2\rho}{\beta E_c}+\Psi\left(\frac{\beta E_c}{2\pi^2 \rho}\right)\right]\, ,
\end{equation}
where $\Psi$ denotes the digamma function and  $c=0.5772\ldots$ Euler's constant. Further, the charging energy is $E_c=2 e^2/C$ and we introduced the dimensionless resistance $\rho=R/R_Q$  with $R_Q=h/4 e^2$. Usually, $\rho\ll 1$ for circuits operated in the overdamped regime.
The classical Smoluchowski range corresponds to the high temperature limit $\eta\hbar\beta\equiv \beta E_c/(\pi\rho)\ll 1$, where $\Lambda\approx \beta E_c/\pi^2\ll 1$, while at low temperatures $\beta E_c/(\pi\rho)\gg 1 $ quantum fluctuations are substantial according to  $\Lambda \approx 2\rho\, {\rm ln}(\beta E_c/\pi^2 \rho)$. The generalization of the classical Langevin equation follows from (\ref{eq:lan2}) by replacing $Z_{Cl}\to Z_Q\equiv \sqrt{\beta \, D_Q(\theta)}\, Z_{Cl}$ which describes a classical stochastic process with multiplicative noise.

In the sequel we consider an atomic point contact with one conduction channel with transmission probability $\tau \in [0,1]$. Generalizations are straightforward. As it is well-known the current through the contact is then carried by two Andreev bound states with energies $E_{\pm}(\theta,\tau)=\pm\Delta \sqrt{1-\tau \sin^{2}(\theta/2)}$ ($\Delta$ is the superconducting gap). If we restrict ourselves to voltages much smaller than $\Delta$ and $k_{\rm B} T$, there are no Landau-Zener transitions between Andreev states and an adiabatic approximation applies. Thus, the
 current-phase relation gets
\begin{equation}
I(\theta,\tau)=\frac{e\Delta}{2\hbar} \frac{\tau \sin(\theta)}{\sqrt{1-\tau \sin^{2}(\theta/2)}}\  {\rm tanh}\left[\frac{\beta E_{+}(\theta,\tau) }{2}\right]\, .
\label{eq:curfase}
\end{equation}
This expression simplifies to the known sinusoidal relation for tunnel junctions in the low transmission limit ($\tau \rightarrow 0$) and is proportional to $\sin(\theta/2)$ in the ballistic limit for $\tau \rightarrow 1$. In this latter domain and for externally driven contacts higher harmonics become relevant such that apart from the conventional
integer Shapiro steps also fractional ones appear. This situation has been studied in the classical realm in Ref.~\cite{dupret}. Here, we focus on the low temperature region where the classical description must be extended to include quantum fluctuations as discussed above.

Before we proceed, let us specify the domain in parameter space where the strong friction approach and the modeling of the environment in terms of an ohmic resistor apply. With respect to the first issue we consider circuits of the type shown in Fig.~\ref{fig:circuit}.
Then, roughly speaking, the friction constant must sufficiently exceed all other relevant frequencies. In the mechanical analog this means $\eta\gg \omega_J^2\hbar\beta, \omega_J^2/\eta, e V_{ac}/\hbar, \omega$ with plasma frequency $\omega_J=\sqrt{E_J E_c}/\hbar$ and Josephson energy $E_J=\Delta (1-\sqrt{1-\tau})$. Note that the last condition $\eta\gg \omega$ enures that the external driving acts on time scales sufficiently larger than the relaxation time for momentum which is of order $1/\eta$. In terms of circuit parameters one has
\begin{equation}
\frac{E_c}{\pi \rho} \gg \pi \rho E_{J}, \beta E_{c} E_{J}, {eV_{ac}}, \hbar\omega\,
\label{constraint}
\end{equation}
with $\rho\ll 1$.
In addition, as discussed above, the ratio $\hbar\beta\eta\equiv\beta E_c/(\pi\rho)$ controls the impact of quantum fluctuations. Further, the adiabatic description (\ref{eq:curfase}) is justified if $\hbar\omega\ll 2\Delta \sqrt{1-\tau}$ to avoid driving induced mixing of the two Andreev surfaces.

With respect to the second issue, the modeling of the environment as being purely ohmic is of course a crude approximation to actual experimental set-ups. Any realistic circuit exhibits at least a cut-off frequency $\Omega_c$ due to unavoidable additional capacitances. The Smoluchowski description remains valid as long as there is still a time scale separation between relaxation in phase [approach of a quasi-stationary state for $P(\theta,t)$] and the response time of the environment, i.e., $\eta/\omega_0^2 \gg 1/\Omega_c$. In fact, it turns out that inertia effects (finite capacitance) and a more refined modeling of the electromagnetic environment lead for sufficiently large friction to only minor deviations from the classical Smoluchowski prediction \cite{Chauvintesis,dupret} (they are relevant for a detailed quantitative analysis of actual circuits though). For the quantum case considered in the sequel, the same is true if $\hbar\beta>1/\Omega_c$ with $\hbar\beta$ being at low temperatures the relevant scale for the coarse graining in time \cite{qmsmolu2}.

\section{Current-voltage characteristics}\label{numerics}

Mean values of relevant observables are determined by the distribution $P(\theta, t)$ determined from the QSE. Since most of the results can only be obtained numerically, we switch in this section to dimensionless quantities and scale energies in units of $\Delta$, frequencies in units of $\Delta/\hbar$, and times in units of $\hbar/(\Delta \rho)$. In particular, this means to measure temperature in units of $\Delta/k_{\rm B}$ and currents in units of $I_c=E_J/\Phi_0$.
The dimensionless QSE then reads
\begin{equation}
\frac{\partial P(\theta,t)}{\partial t}=-\dfrac{\partial }{\partial \theta}\left[\frac{\partial U(\theta,t)}{\partial \theta}P(\theta,t) + \frac{\partial D_{Q}(\theta) P(\theta,t)}{\partial \theta}\right]\, ,
\label{eq:curmean2}
\end{equation}
which is in fact a continuity equation for the probability, i.e., $\partial P/\partial t+\partial J/\partial\theta=0$ with
the probability flux
\begin{equation}
J(\theta,t) \equiv \frac{\partial U(\theta,t)}{\partial \theta}P(\theta,t) +  \frac{\partial D_{Q}(\theta) P(\theta,t)}{\partial \theta}\, .
\label{eq:curmean2b}
\end{equation}

Now, these expressions determine mean values with respect to phase $\overline{(...)}$ and time $<...>$. For the current one has
\begin{equation}
\overline{\langle I(\theta)\rangle}=\int_{0}^{2\pi} d\theta  \int_{-\infty}^{\infty} dt I(\theta) P(\theta,t)
\end{equation}
and the voltage across the contact $\overline{\langle V\rangle}=\overline{\langle \dot{\theta}\rangle}$ follows as
\begin{equation}
\overline{\langle V\rangle}=\int_{0}^{2\pi} d\theta  \int_{-\infty}^{\infty} dt J(\theta,t)\, .
\end{equation}
Due to the periodicity of the potential $U(\theta,t)$ in phase and time [cf.~Eq.~(\ref{eq:pot})], one expands density and current according to
\begin{equation}
P(\theta,t)= \sum_{n,k \in Z} P_{n,k} e^{i k \theta+i n \omega t}
\end{equation}
\begin{equation}
J(\theta,t)= \sum_{n,k \in Z} J_{n,k} e^{i k \theta+i n \omega t}
\end{equation}
with the normalization condition
\begin{equation}
P_{n,0} = \delta_{n,0}/2\pi\, .
\end{equation}
Further, one writes due to (\ref{eq:curfase})
\begin{equation}
I({\theta}) = \sum_{m=1}^{\infty} I_{m}(\theta,\tau) \sin(m \theta)
\label{eq:curexp}
\end{equation}
as well as
\begin{equation}
D_{Q}({\theta}) = \sum_{m=0}^{\infty} D_{m}(\theta,\tau) \cos(m \theta)\, .
\label{eq:difexp}
\end{equation}

This way, the QSE (\ref{eq:curmean2}) is cast in an algebraic equation for the expansion coefficients, namely,
\begin{eqnarray}
\frac{n}{k} \omega P_{n,k} &=& I_{b}P_{n,k}-i \frac{V_{ac}}{2\rho}(P_{n-1,k}+P_{n+1,k}) \nonumber\\
&&+ \sum_{m=1}^{\infty} I_{m}(P_{n,k-m}-P_{n,k+m})   \nonumber\\
&&+ i k \sum_{m'=0}^{\infty}D_{m'}(P_{n,k-m'}+P_{n,k+m'})
\label{seteq}
\end{eqnarray}
Practically, one works on a two-dimensional grid for $n, k$ with $|n|\leq N_{\rm max}, |k|\leq K_{\rm max}$.
As already pointed out in \cite{dupret} the corresponding set of $N_{\rm max} \times K_{\rm max}$ coupled equations can be associated with a non-Hermitian lattice model for particles on a square lattice. In particular, one observes that there is a coupling between chains $n$ and $n\pm 1$ proportional to $V_{ac}$, a coupling between chains $k$ and $k\pm m$ proportional to the $m$-th harmonic of the Josephson current, and a coupling between chains $k$ and $k\pm m'$ proportional to the $m'$-th harmonic of the quantum diffusion coefficient. Note that this latter coupling is absent in the classical regime.

Now, the orthogonality of circular functions allows to express the mean current and voltage as
\begin{equation}
\overline{<I(\theta)>}=\sum_{k\in Z} P_{0,k} I_{-k}
\label{curr}
\end{equation}
\begin{equation}
\overline{<V>}=\rho\ \left(I_{b}-\sum_{k\in Z} P_{0,k} I_{-k}\right)
\label{volt}
\end{equation}
meaning that we only need to calculate $P_{0,k}$ explicitly. In fact, peaks in this probability coefficient are related to the observed Shapiro steps of order $n/k$ in the $I-V$ characteristics.

To solve  (\ref{seteq}) numerically we define vectors $\overrightarrow{P_{n}}\equiv (\ldots P_{n,k},\ldots,
P_{n,1},P_{n,-1},\ldots,P_{n,-k},\ldots)$ and
$\overrightarrow{I}\equiv (\ldots I_{k},\ldots,I_{1},I_{-1},\ldots,I_{-k},\ldots)$ and matrices
\begin{eqnarray}
(L_{n})_{k,k'} &\equiv &\left(\frac{n}{k} \omega P_{n,k}+ I_{b}\right)\delta_{k,k'} \nonumber\\
&&- I_{m}\left(\delta_{k',k-m}-\delta_{k',k+m}\right)\nonumber\\
&& -i k D_{m'}\left(\delta_{k',k-m'}+\delta_{k',k+m'}\right)\,
\label{setmat}
\end{eqnarray}
so that (\ref{seteq}) takes the compact form
\begin{equation}
L_{n} \overrightarrow{P}_{n}= \frac{V_{ac}}{2\rho}(\overrightarrow{P}_{n-1}+\overrightarrow{P}_{n+1})
+\delta_{n,0} \overrightarrow{I}\, .
\label{setmat2}
\end{equation}
This equation is solved via a recursive procedure (continued fraction method-upward iteration) by introducing for $n>0$ the auxiliary quantity $S_{n+1} \overrightarrow{P_{n}}= \frac{V_{ac}}{2\rho}(\overrightarrow{P}_{n+1})$ and for $n<0$ by defining with $\overline{n}\equiv-n$ the quantity $
S_{\overline{n}+1} \overrightarrow{P}_{\overline{n}}= \frac{V_{ac}}{2\rho}(\overrightarrow{P}_{\overline{n}+1})$. Accordingly, one has a simple equation for the relevant probability coefficients $\overrightarrow{P}_{0}= [L_{0}- S_{1} -S_{\overline{1}}]+\overrightarrow{I}$,
where
\begin{equation}
\begin{split}
S_{1(\overline{1})}&= -
\dfrac{\mu}{L_{1(\overline{1})}-\dfrac{2\mu}{L_{2(\overline{1})}-\dfrac{\mu}{L_{3(\overline{3})}-...} }}
\end{split}\,
\label{setmat6}
\end{equation}
with the abbreviation $\mu=(V_{ac}/2 \rho)^2$.

\section{Results}\label{results}

We start with a brief discussion of actual experimental parameters and proceed with a presentation of the numerical results.

\subsection{Approximations and parameters ranges}

We take typical experimental values for atomic contacts with Al electrodes \cite{stein} with a superconducting gap $\Delta \simeq 180 \mu eV$. Temperatures are varied between about $10$mK and $100$mK and the circuit is assumed to have an ohmic resistance of $200\Omega$ such that $\rho \ll 1$ and a capacitance on the order of fF.
Typical microwave frequencies are $\hbar\omega \sim 10^{-2} \Delta - 10^{0} \Delta$ with $\omega >\rho \Delta/\hbar$ to observe fractional Shapiro steps.
Within this range of parameters the conditions in (\ref{constraint}) are fulfilled and phase diffusion is supposed to be affected by quantum fluctuations in the strong damping regime. In particular, $\beta E_c/\pi\rho\gg 1$ so that the energy scale related to friction $\hbar\eta$ by far exceeds the thermal energy scale $k_{\rm B} T$.

We now solve numerically the recursion (\ref{setmat6}) where convergence depends on the maximum number of spatial and temporal harmonics, the temperature and the external voltage. The numerics is quite sensitive at low temperatures and high voltages, but in the chosen ranges of these parameters  accurate data can be achieved with $N_{\rm max}=90$ and $K_{\rm max}=45$.

\subsection{Numerical Results}

To analyze the Shapiro step structure we calculate numerically, from (\ref{curr}) and (\ref{volt}), mean currents $\overline{\langle I(\theta)\rangle}$ and mean voltages $\overline{\langle V\rangle}$ (in the figures $I$ and $V$ respectively), over the temperature range specified above and for various transmission coefficients.
\begin{figure}
\begin{center}
\epsfig{file=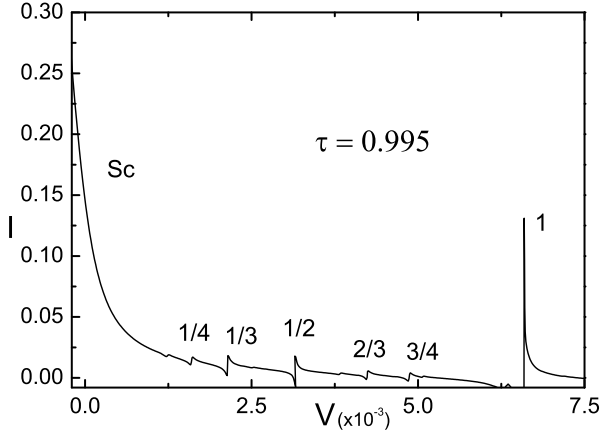,width=10cm}
\end{center}
\caption{$I-V$ curves for an ac-driven atomic point contact with $\tau=0.995$  and $T=0.005$.  Other parameters are $\omega=2\pi.10^{-3}$, $\eta=5$, $V_{ac}=5.10^{-3}$ , $R=10^{-3}$  (dimensionless units, see beginning of Sec.\ref{numerics}).}
\label{fig:IVTotal}
\end{figure}
Figure~\ref{fig:IVTotal} shows a typical $I-V$ curve with the first integer and several fractional resonances in the quantum regime (low temperatures) and for a high transmissive junction.
\begin{figure}
\epsfig{file=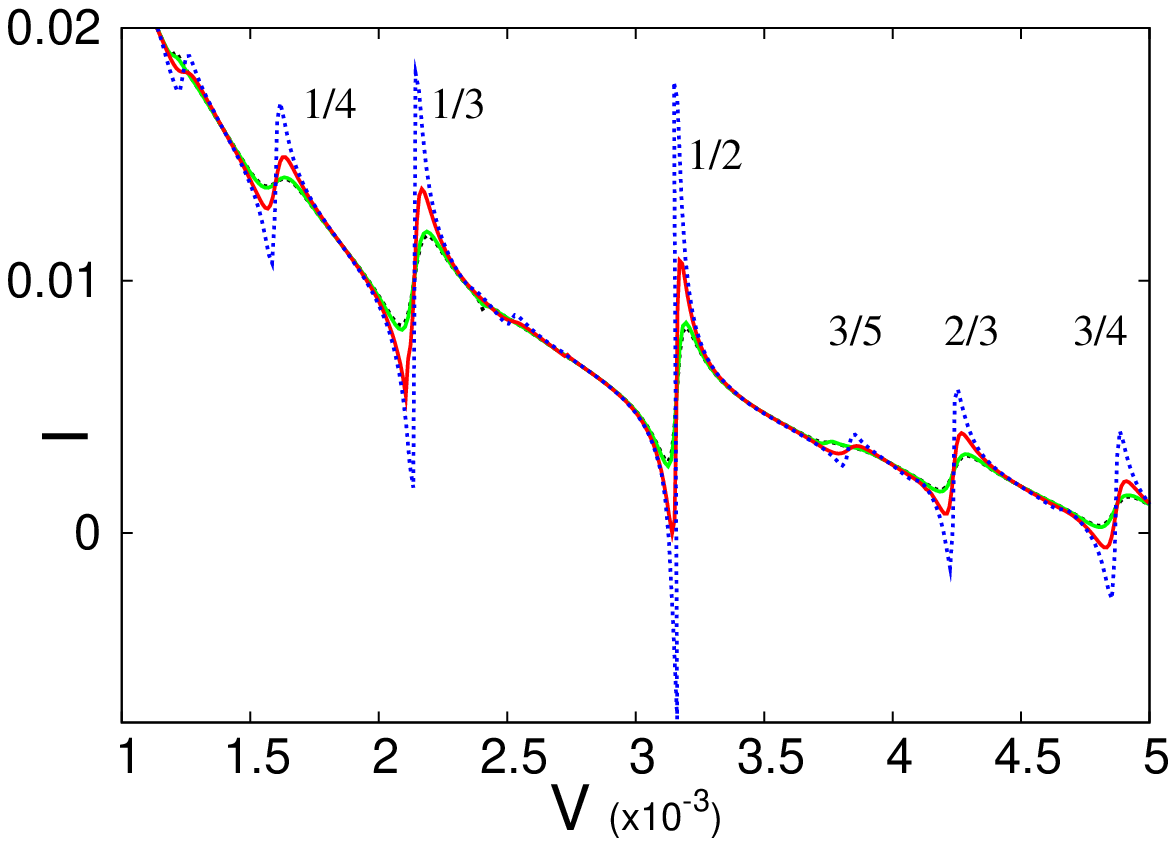,width=8.75cm}
\epsfig{file=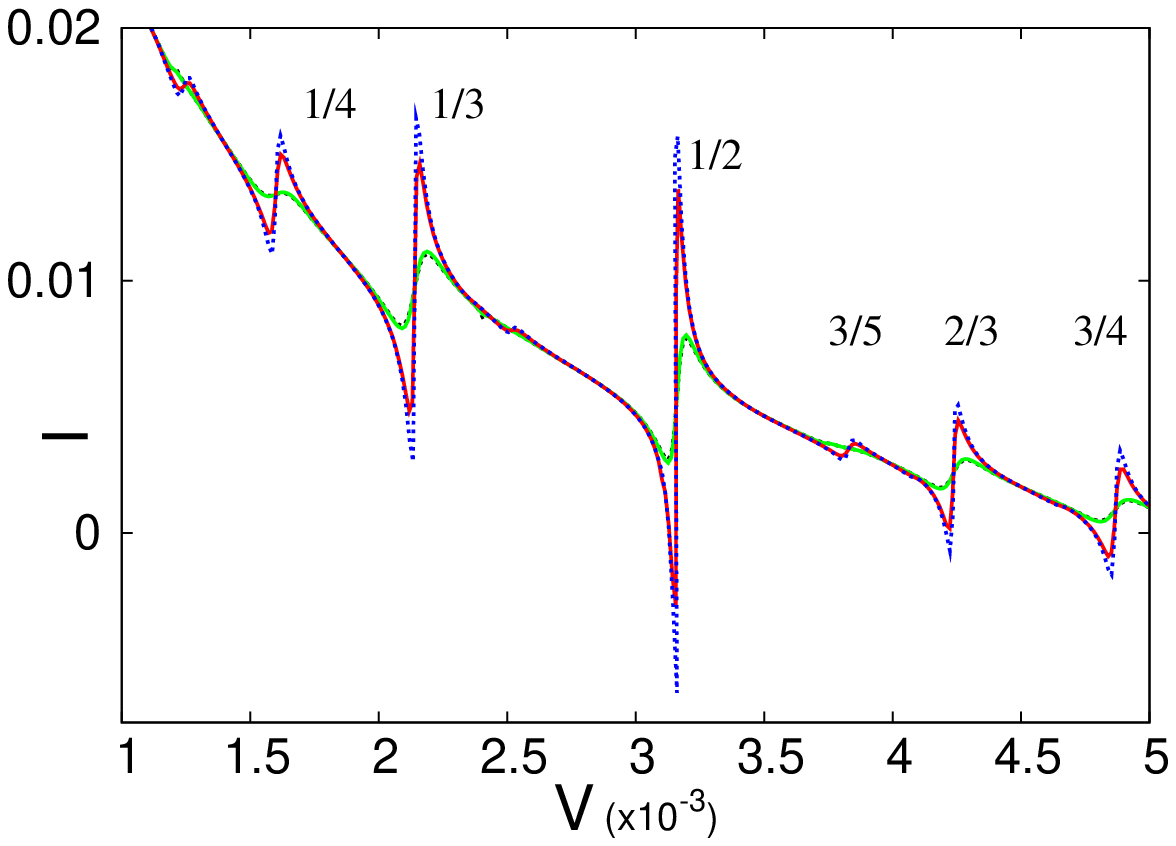,width=8.75cm}
\caption{Fractional Shapiro steps for a high ($\tau=0.995$, top panel) and lower ($\tau=0.99$, bottom panel) transmissive channels. Red(green)-solid lines depict data obtained with the QSE at $T=0.006(0.02)$, while blue(black)-dotted lines describe results from the SE at $T=0.006(0.02)$. At a higher temperature $T=0.02$ both approaches give identical curves. Other parameters are as in Fig.~\ref{fig:IVTotal}.}
\label{fig:IvsVzoomComp}
\end{figure}
While resonances appear in a pattern very similar to the known classical ones,
the role of quantum fluctuations is revealed when one compares low temperature results obtained with the SE and those gained with the QSE, respectively (Fig.~\ref{fig:IvsVzoomComp}). At higher temperatures the diffusion coefficient $D_Q\to D_{Cl}$ such that both descriptions deliver identical data, but there are substantial deviations at low temperatures where $D_Q(\theta)>D_{Cl}$. Indeed, the classical equation predicts much sharper resonances than the quantum one, where the reduction in height and the increase in width is more striking at higher transmissions. This smearing out is basically absent away from the resonances.
\begin{figure}[h]
\epsfig{file=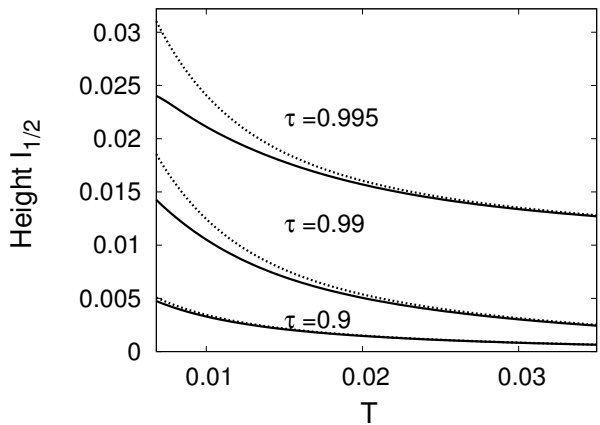, width=8.5cm}
\caption{Resonant peak height vs. temperature for the Shapiro step 1/2 according to the classical approach (dotted) and the quantum one (solid) and transmission coefficients $\tau=0.995$ (shifted upwards by 0.01), $\tau=0.99$, and $\tau=0.9$. Other parameters are the same than in previous figures.}
\label{fig:12Temp}
\end{figure}

\begin{figure}[h]
\begin{center}
\epsfig{file=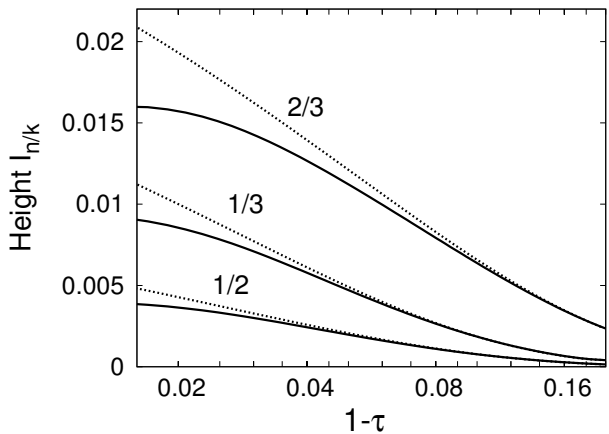, width=8.5cm}
\end{center}
\caption{Peak heights $I_{n/k}$ vs. 1-$\tau$ for resonances with $n/k$=1/2, 1/3 and 2/3 (from top to bottom) at temperature $T=0.006$. Solid lines correspond to the quantum case and dotted lines to the classical one. Parameters are the same than in previous figures.}
\label{fig:TotalTP}
\end{figure}

In order to have a better insight in this behavior the heights of the fractional peak $I_{1/2}$ [calculated from $(I_{\rm 1/2, max}-I_{\rm 1/2, min})/2$ as differences between maximal and minimal peak values] are plotted as functions of temperature in Fig. \ref{fig:12Temp}. As already discussed, quantum fluctuations reduce the peak heights at lower temperatures and for higher transmissive channels. Since the overall peak structure is not altered,
one may misleadingly describe a reduced height within the classical approach by an effectively enhanced temperature. However, while it is true that the quantum diffusion coefficient is typically larger than the classical one ($D_Q>D_{Cl}\equiv k_{\rm B} T$), due to its dependence on the phase $\theta$ the QSE can in general not simply be reduced to the SE by replacing $T$ by an effective temperature (see next section).
Variations of peak heights with increasing transmission are illustrated in Fig.~\ref{fig:TotalTP} for several fractional resonances. Interestingly, mean values $I_{n/k}$ saturate in the quantum case towards the ballistic limit $\tau\to 1$ with larger deviations from the classical data for higher order steps.

\begin{figure}
\epsfig{file=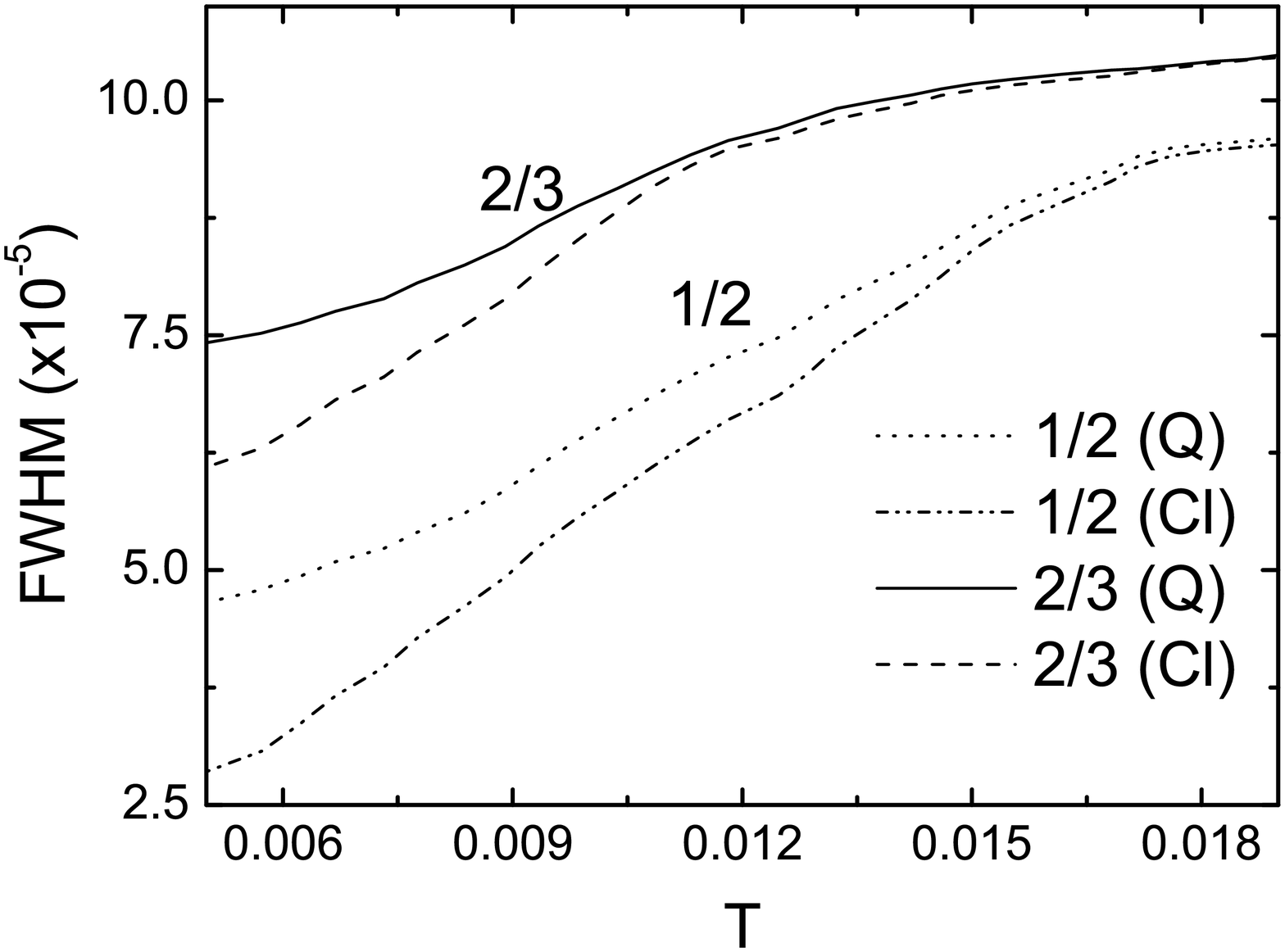, width=9.5cm}
\epsfig{file=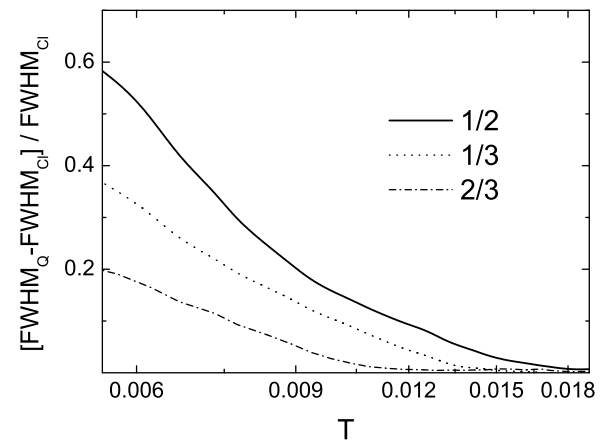, width=9.5cm}
\caption{$FWHM$ vs. $T$ for $\tau = 0.995$ and various fractional resonances (top) together with the relative strength of quantum fluctuations
$(FWHM_{Q}-FWHM_{Cl})/FWHM_{Cl}$ (bottom). Other parameters are the same as in previous figures.}
\label{fig:FWMH}
\end{figure}

Apart from reduced heights quantum effects appear as a widening of the resonances. A natural magnitude to quantify this, is the full width at half maximun ($FWHM$) of the  peaks. As expected, we see in Fig.~\ref{fig:FWMH} that the spreading of the quantum peaks exceeds that of the classical ones at lower temperatures with the $FWHM$ taking larger values for higher harmonics. Both predictions coincide only at relatively elevated temperatures. The relative strength of quantum fluctuations is larger at lower fractional steps, cf.~Fig.~\ref{fig:FWMH}.

The crucial question is, of course, whether the influence of quantum fluctuations seen above could actually be detected in a real experimental set-up. In fact, for contacts with different transmissions also experimental data (see e.g.\ Ref.\cite{Chauvintesis}) deviate from results of the adiabatic classical theory. There are at least three possible explanations for this discrepancy: Landau-Zener (LZ) transitions between adiabatic surfaces, charging effects and associated quantum fluctuations, and spurious noise. With respect to the first one,  it was shown  in Ref. \cite{fritz} that nonadiabatic LZ transitions enhance the magnitude of the supercurrent peak in almost ballistic channels. This effect is stronger at slightly elevated temperatures and in highly transmissive channels.
Physically, nonadiabatic transitions between $E_-$ and $E_+$ surfaces only occur if the diffusive passage of the phase through the LZ-range around $\theta=\pi$ is sufficiently fast compared to the instantaneous relaxation time of momentum. Accordingly, LZ-transitions are suppressed towards very low temperatures
in contrast to what is observed for quantum fluctuations (see previous figures). Hence, we are left with either quantum fluctuations or spurious noise or, what is most likely, both to explain the addressed differences. Spurious noise alone can be captured by an effective temperature which is indeed the strategy that has been followed in \cite{Chauvintesis}. For this purpose, we analyze in the following to what extent this concept may also be applicable to effectively include quantum noise.

\section{Quantum diffusion and effective temperature}\label{effective}

To describe the dynamics of the phase in the vicinity of a resonance, we consider instead of the current biased circuit in Fig.~\ref{fig:circuit} (Norton representation) the completely equivalent circuit where the contact is voltage biased (Thevenin representation). Within the adiabatic approximation and in the overdamped quantum range $\beta E_c/\pi\rho\gg 1$  the voltage across the contact  is then given by (again in physical dimensions)
\begin{equation}
V\equiv \Phi_0\dot{\theta}(t) =  R \, I(\theta(t))+ V_{b}+V_{ac} \cos(\omega t)+ R\, {Z}_{Q}(\theta(t))
\label{eq:lan4}
\end{equation}
with the voltage bias $V_b=I_b R$ and the quantum noise $Z_{Q}(\theta)=\sqrt{\beta\, D_{Q}(\theta)}\, Z_{Cl}$.

For a perfect voltage bias (no noise) and in absence of external driving the phase evolves as $\theta(t)=\theta(0)+\omega_0 t$ with the
Josephson frequency $\omega_0=2 e V_b/\hbar$. In presence of an ac-drive the $n/k$-Shapiro resonance  appears if $k\omega_0=n \omega$ at a corresponding voltage $V_b=\langle{V}\rangle=(n/k) \Phi_0\omega$. Thus, right at the center of the $n/k$ resonance no dc-current flows and according to (\ref{eq:lan4})  the diffusive motion of the phase can be expressed as
\begin{equation}
\theta(t) = 2\nu \sin(\omega t)+\frac{n}{k} \omega t + \delta(t)
\label{eq:fase2}
\end{equation}
with $\delta$ being the stochastic component of the phase on top of the dominating deterministic part $\theta_0(t)=2\nu \sin(\omega t)+\frac{n}{k} \omega t$ where $\nu=e V_{ac}/\hbar\omega$. Upon inserting this expression in (\ref{eq:lan4}) one arrives at
\begin{equation}
\Phi_0 \dot{\delta}(t) = R I(\theta(t))+V_{b}-\frac{n}{k} \Phi_0\omega  + Z_{Q}(\theta(t))\, .
\label{eq:fase2b}
\end{equation}
Apparently, the dynamics of the stochastic part $\delta$ is much slower than that of the deterministic part since $|V_{b}-(n/k) \omega|\ll (n/k)\omega$ and $I(\theta)\approx 0$. This separation of time scales can be exploited when calculating time averaged currents.
Namely, plugging the result (\ref{eq:fase2}) into the Fourier expansion (\ref{eq:curexp}) of the current leads first to
\begin{eqnarray}
I({\theta}(t)) &=  &\sum_{m=1}^{\infty} I_{m}(\tau)\Biggl\{\sin[m\alpha_{nk}(t)] \Bigl[J_0(2m\nu)\nonumber\\
&& +2 \sum_{p\geq 1}J_{2p}(2m\nu) \cos(2p\omega t)\Bigr]+2 \cos[m \alpha_{nk}(t)]  \nonumber\\
&& \times \sum_{p\geq 1}J_{2p+1}(2m\nu)\cos[(2p+1)\omega t]\Biggr\}
\label{eq:fase5}
\end{eqnarray}
with the abbreviation $\alpha_{nk}(t)=(n/k)\omega t+\delta(t)$ and the Bessel function $J_p$.
Taking now time averages over one period of the external drive and accounting for the time scale separation between deterministic and stochastic dynamics of the phase we obtain
\begin{eqnarray}
I({\theta}(t))&\approx &2 \sum_{m \geq 1} I_{m}(\tau)\, \sin[m \delta(t)] \nonumber\\
&&\times \sum_{p \geq 1} J_{2p}(2m\nu) \langle \cos\left(\frac{m n}{k}\omega t\right) \cos(2p\omega t)\rangle\nonumber\\
&&- 2 \sum_{m \geq 1} I_{m}(\tau)\, \sin[m \delta(t)] \nonumber\\
&& \times \sum_{p \geq 0} J_{2p}(2m\nu) \langle \sin\left(\frac{m n}{k}\omega t\right) \sin(2p\omega t)\rangle
\label{eq:fase6}
\end{eqnarray}
Here, the averages can only take two values, namely, 0 or $1/2$ depending on the relation between $\frac{m n}{k},2p$ and $(2p+1)$. Eventually, this yields the current phase relation in the vicinity of the $n/k$ Shapiro resonance
\begin{equation}
I({\theta}(t)) \approx  \sum_{l \geq 1} (-1)^{ln} I_{lk}(\tau) J_{ln}(2lk\nu) \sin[l\phi_k(t)]\, ,
\label{eq:fase7}
\end{equation}
where we put $m=l\, k$, $2p=l\, n$ with $l\in N$ and introduced the scaled phase $\phi_k(t)=k\,\delta(t)$.

The expression (\ref{eq:fase7}) is then inserted into (\ref{eq:fase2b}) to provide an approximate equation of motion for $\phi_k$.
While in general solutions are  accessible only numerically, insight is already gained by keeping just the term with $l=1$ in (\ref{eq:fase7}), i.e.,
\begin{eqnarray}
\Phi_0 \dot{\phi_k}& =& k\left(V_{b}-\frac{n}{k}\Phi_0 \omega\right)+R\, (-1)^n k I_k J_n(2 k \nu) \sin(\phi_k) \nonumber\\
&& + k\, R\, Z_{Q}(\theta_0+\phi_k/k)\, .
\label{eq:faseeff}
\end{eqnarray}
 In contrast to the classical case, here the noise term
 also depends on the phase.
 In the classical regime, one shows that (\ref{eq:faseeff}) is identical to the equation of motion for the phase in absence of ac-driving if
 parameters are renormalized \cite{Chauvintesis}: $I_{c, \rm eff}=|k I_k  J_n(2 k \nu)|$, $T_{\rm eff}=k^2 T$, $V_{b,\rm eff}=k (V_{b}-\frac{n}{k}\Phi_0 \omega)$.
 This way, one gains replicas of the Ivanchenko-Zil'berman expression for the dc-supercurrent around each $n/k$ peak, namely,
 \begin{equation}
\langle I \rangle(V_{b}) = I_{c, \rm eff} f_{IZ}\left(\dfrac{V_{b, \rm eff}}{I_{c, \rm eff}} , \dfrac{I_{c, \rm eff}}{T_{\rm eff}}\right)
\label{eq:ivan}
\end{equation}
with $f_{IZ}(x,y)={\rm Im}\{I_{1-i x y}(y)/I_{-i x y }(y)\}$ the modified Bessel function of first kind.
 Now, in the quantum regime in leading order we may put $Z_Q(\theta)\approx Z_Q(\theta_0)=\sqrt{\beta D_Q(\theta_0)} \, Z_{Cl}$
 such that a similar renormalization applies, however, with a modified temperature scaling, i.e.,
 \begin{equation}
 T_{q, \rm eff}= k^2 \langle \beta D_Q(\theta_0(t))\rangle \, T > T_{\rm eff}\, .
 \label{effectiveT}
 \end{equation}
 This effective temperature depends also on the dissipation strength, the driving frequency and amplitude, and is a nonlinear function of the actual environmental temperature. We note that the actual experimental data \cite{Chauvintesis} do not follow the scaling of $T_{\rm eff}$ with $k$, but rather can only be described by a much higher effective temperature $T_{\rm eff, exp}$ which even affects the integer Shapiro resonances and has been attributed to spurious noise in the circuitry. The above enhancement due to quantum fluctuations may partially contribute to $T_{\rm eff, exp}$, however, is not able to completely account for the discrepancy between $T_{\rm eff, exp}$ and $T_{\rm eff}$.

 Beyond the case for $l=1$ progress is achieved by assuming $|\Lambda \beta U''(\theta)|\ll 1$ so that one may expand [cf.~Eq.~(\ref{eq:qsmol})] $\beta D_Q(\theta)\approx 1+\Lambda \beta I'(\theta)$ with
 \begin{equation}
I'({\theta})\approx  \sum_{l \geq 1} l\, q (-1)^{ln} I_{lk}(\tau) J_{ln}(2lk\nu) \cos(l\phi_k)\, .
\label{eq:pot3}
\end{equation}
Thus, if contributions with sufficiently large $l$ are relevant, one may no longer replace $\cos(l\phi_k)\to 1$ meaning that quantum noise {\em cannot} be captured by a global effective temperature. Instead, its phase dependence leads to a local ''temperature'' and $n/k$ resonances are {\em not} simply replicas of the supercurrent peak. To extract signatures of this breakdown of the universal scaling behavior (\ref{effectiveT}), the residual spurious noise dominating $T_{\rm eff, exp}$ must be substantially reduced.
 We note that the findings of Grabert et. al \cite{grabert} who studied the supercurrent phase diffusion in absence of driving in the {\it tunnel limit} and at low temperatures within a time-dependent perturbation theory, obtained also an extension of the Ivanchenko Zil'berman expression similar to (\ref{eq:ivan}) but with an effective Josephson energy. This result was later reproduced within the QSE-formulation \cite{Anker}.

\section{Summary}\label{conclusions}

In summary, the impact of quantum fluctuations on fractional Shapiro resonances, a hallmark of a non-sinusoidal current phase relations, is analyzed for atomic point contacts with highly transmitting channels in the presence of microwave.
Known experimental $I-V$ results \cite{Chauvin,Chauvintesis} exhibit substantial deviations when compared with predictions from a classical adiabatic theory. While one explanation has been the appearance of spurious noise in the circuit, here, we find that quantum fluctuations may give rise to a similar effect. Departures from the classical approach become relevant for highly transmissive channels and for sufficiently low temperatures. An effective description of quantum noise in terms of an effective temperature only applies to
contacts with almost sinusodial current-phase relations which may offer a way to distinguish between classical noise and quantum fluctuations in high transmissive contacts. As a prerequisite, however, unspecific spurious noise sources in the circuitry must be under control so that contacts are embedded in heat baths with temperatures of about $T\sim 40$ mK or below.

The results that we have obtained correspond to Al point contacts with a superconducting gap of $ \Delta_{\rm Al} \sim 200 \mu eV$. For this metal our model predicts that quantum fluctuations play a pronounced role for very low temperatures ($T < 40$ mK). Landau-Zener transitions are negligible if  $T \ll T_{\rm LZ}\sim 0.5 \Delta_{\rm Al}/k_{\rm B} \approx 1$ K (for transmissions $\tau>1-10^{-4}$) \cite{fritz}. Experimentally, signatures of quantum fluctuations are expected to be even more dominant for materials with larger superconducting gaps such as e.g.\  Nb with $\Delta_{\rm Nb} \approx 3600 \mu eV$ leading to a clear separation between $T_{\rm LZ} \approx 9$ K and the temperature range where typical experiments are performed (between 30 mK and 150 mK).

\begin{acknowledgements}
We acknowledge stimulating discussion with A. Levy Yeyati. This work was supported by PIP Conicet (FC) and by the DFG through SFB569 (JA).
\end{acknowledgements}

\end{document}